\documentclass[twocolumn,showpacs,preprintnumbers]{revtex4}
\usepackage{amssymb}
\usepackage{graphicx}

\input{tcilatex}

\begin{document}

\title{Exciton effects in a scaling theory of intermediate valence and Kondo
systems}
\author{V. Yu. Irkhin and M. I. Katsnelson}
\address{Institute of Metal Physics, Ekaterinburg 620219, Russia\\
Department of Physics, Uppsala University, Uppsala 751 21 Sweden}

\begin{abstract}
An interplay of the Kondo scattering and exciton effects ($d-f$ Coulomb
interaction) in the intermediate valence systems and Kondo lattices is
demonstrated to lead to an essential change of the scaling behavior in
comparison with the standard Anderson model. In particular, a marginal
regime can occur where characteristic fluctuation rate is proportional to
flow cutoff parameter. In this regime the ``Kondo temperature'' itself is
strongly temperature dependent which may give a key to the interpretation of
controversial experimental data for heavy fermion and related systems.
\end{abstract}

\maketitle

\address{Institute of Metal Physics, Ekaterinburg 620219, Russia\\
Department of Physics, Uppsala University, Uppsala 751 21 Sweden}

There is an interesting class of rare-earth compounds such as Ce$_4$Bi$_3$Pt$%
_3$, SmB$_6$, SmS under pressure (the ``golden'' phase), TmSe, YbB$_{12}$
which were called earlier intermediate valence (IV) compounds and now are
treated as ``heavy-fermion (HF) semiconductors'', or ``Kondo insulators''
(for a review, see Refs.\onlinecite{rise1,rise2}). Various names emphasize
different peculiarities of these compounds. As for electron energy spectrum,
most of them are narrow gap semiconductors with an anomalously small energy
scale (the gap width), of the order of tens or hundreds of Kelvins (see Ref.%
\onlinecite{rise2} and a review of earlier experiments in Ref. %
\onlinecite{IK86}). At the same time, they do demonstrate intermediate
valence of the rare earth ions (usually between 2+ and 3+) in a number of
properties, e.g., lattice constants (which are intermediate between those
for isostructural compounds with di- and trivalent ions), core level spectra
(which are a mixture of the spectra of di- and trivalent ions with
comparable weights), and many others \cite{rise1,conf1,conf2}.

As well as for the HF metals, the origin of this small energy scale is a key
point to understand anomalous properties of the IV compounds. For the HF
metals it is commonly accepted now that they are the Kondo lattices, which
means that this energy scale (the Kondo temperature $T_K$) is a width of the
Kondo resonance owing to spin-dependent scattering of conduction electrons
by $f$-electron centers \cite{hewson}. As a result of an interplay of the
Kondo effect and interatomic magnetic interactions, the $T_K$ value for a
lattice can strongly differ from that for an isolated impurity \cite{IK97},
spin fluctuations being of crucial importance for the HF behavior. It is
very natural to expect that these effects are important also for the IV
compounds. At the same time, valence or \textit{charge} fluctuations should
be also considered. They are determined in part by the Coulomb
(``Falicov-Kimball'') interaction between conduction and localized electrons
\cite{falicov}. Taking into account these interactions together with the
hybridization processes it is possible to describe the IV state as a kind of
exciton condensation \cite{IK86,stevens}. Note that in the IV regime the
spinless Falicov-Kimball model with hybridization is formally equivalent to
the anisotropic Kondo problem, different valent states playing the role of
pseudospin ``up'' and ``down'' states \cite{schlottm}. It is the degeneracy
of quantum states for a scattering center which is important for the
formation of the Kondo resonance \cite{cox}. In the IV case the divalent and
trivalent states are degenerate by definition, so that this analogy is not
surprising. Therefore it is natural to consider the formation of the Kondo
resonance for the IV compounds taking into account both spin and charge
fluctuations, or, equvalently, both the ``Kondo'' and exciton
(``Falicov-Kimball'' ) effects. This is the aim of the present work. Since
there is no clear demarcation between the IV and Kondo systems, it will be
shown that the excitonic effects may be relevant also for the latter case.

To investigate effects of interaction of current carriers with local moments
we use the Hamiltonian of the asymmetric infinite-$U$ $SU(N)$ Anderson model
with inclusion of the Falicov-Kimball interaction (on-site $d-f$ Coulomb
repulsion $G$),
\begin{eqnarray}
\mathcal{H} &=&\sum_{\mathbf{k}m}[t_{\mathbf{k}}c_{\mathbf{k}m}^{\dagger }c_{%
\mathbf{k}m}+V\left( c_{\mathbf{k}m}^{\dagger }f_{\mathbf{k}m}+f_{\mathbf{k}%
m}^{\dagger }c_{\mathbf{k}m}\right)   \nonumber \\
&&+E_{f}f_{\mathbf{k}m}^{\dagger }f_{\mathbf{k}m}]+G\sum_{imm^{\prime
}}f_{im}^{\dagger }f_{im}c_{im^{\prime }}^{\dagger }c_{im^{\prime }}
\end{eqnarray}%
where the on-site $f-f$ Coulomb interaction is put to infinity, so that
doubly occupied states are suppressed, $f_{im}^{\dagger }=|im\rangle \langle
i0|$ are the Hubbard operator ($|im\rangle $ and $|i0\rangle $ are
single-occupied and empty states), we neglect for simplicity the $\mathbf{k}$%
-dependence of the hybridization $V$. Note that similar calculations can be
performed for realistic rare-earth ions, including the case of two magnetic
configurations, see Ref. \onlinecite{ii}.

Following to Ref. \onlinecite{schlottm}, we treat the coherent and
incoherent cases. In the first case dispersion in the spectrum of $f$%
-electrons occurs. For simplicity this is supposed to be proportional to the
conduction electron spectrum, $E_{f\mathbf{k}}=E_{f}+\lambda \varepsilon _{%
\mathbf{k}},\varepsilon _{\mathbf{k}}\varpropto t_{\mathbf{k}}$, $\lambda =-1
$ (the $f$-band has a hole character). In the incoherent regime $\lambda =0,$
so that the $f$-electrons remain localized. Note that in the presence of the
energy gap we always deal with the coherent regime.

The renormalization of the Coulomb parameter $G$ and the hybridization $V$
is obtained, similar to Ref. \onlinecite{schlottm}, from the two-particle
Green's function
\begin{equation}
F_{mm^{\prime }}^\sigma (E)=\langle \langle f_{im}^{\dagger }c_{im^{\prime
}}|c_{im^{\prime }}^{\dagger }f_{im}\rangle \rangle _E
\end{equation}
which determines the vertex. We obtain the singular correction with the
structure
\begin{equation}
\delta F_{mm^{\prime }}^\sigma (E)=GF_{mm^{\prime }}^\sigma (E)\sum_{\mathbf{%
q}}\frac{n_{\mathbf{q}m^{\prime }}}{E-t_{\mathbf{q}}+E_f}  \label{ver}
\end{equation}
where $n_{\mathbf{k}m}=\langle c_{\mathbf{k}m}^{\dagger }c_{\mathbf{k}%
m}\rangle $ is the Fermi function. In the coherent regime, a similar
correction occurs from the dispersion of $f$-states. The correction (\ref%
{ver}) contains a logarithmic Kondo-like divergence owing to charge
fluctuations, which is cut at $E_f$ (the latter quantity plays the role of
the external field in the equivalent anisotropic Kondo model). Unlike the
renormalization of $E_f,$ the renormalizations of $V$ and $G$ do not contain
the degeneracy factor of $N$.

The renormalization of $E_f$ owing to spin-flip processes is obtained in the
second order in hybridization (cf. \cite{haldane,IKZ,ii})
\begin{equation}
\delta E_{fm}=V^2\sum_{m^{\prime }\neq m,\mathbf{q}}\frac{n_{\mathbf{q}%
m^{\prime }}}{E_F-t_{\mathbf{q}}}+n\delta G  \label{def}
\end{equation}
We have taken into account in Eq.(\ref{def}) the Hartree renormalization of $%
f$-level energy, which occurs in the coherent case, $E_f\rightarrow E_f+Gn,$
$n$ being the concentration of conduction electrons; we put in numerical
calculations $n=1,$ which corresponds to the Kondo regime.

To derive the scaling equations for the effective model parameters we use
the poor-man scaling approach \cite{And}. Picking out in the integrals with
the Fermi functions (\ref{ver}), (\ref{def}) the contributions from the
energy layer $C<E<C+\delta C$ near the Fermi level $E_F=0$ and replacing $%
E_f\rightarrow E_f(C),V\rightarrow V(C),G\rightarrow G(C)\ $ we obtain (cf.
Ref. \onlinecite{schlottm})

\begin{eqnarray}
\frac{\partial E_f(C)}{\partial \ln C} &=&-\rho (N-1)V^2(C)+n\frac{\partial
G(C)}{\partial \ln |C|}  \label{gll} \\
\frac{\partial V(C)}{\partial \ln |C-E_f|} &=&-\frac{1-\lambda }{1+|\lambda |%
}\rho V(C)G(C)  \label{wll} \\
\frac{\partial G(C)}{\partial \ln |C-E_f|} &=&\frac{2\lambda }{1+|\lambda |}%
\rho G^2(C)  \label{zl}
\end{eqnarray}
where $\rho $ is the bare conduction-electron density of states at $E_F$.
Earlier \cite{IK86} we have considered the exciton effects with neglecting
spin fluctuations. We will see that the renormalization (\ref{def}) results
in new essential effects.

We have from Eqs.(\ref{wll}) and (\ref{zl})
\begin{equation}
\frac{G(C)}{G(0)}=\left[ \frac{V(C)}{V(0)}\right] ^\alpha ,\alpha =-\frac{%
2\lambda }{1+|\lambda |}  \label{gv}
\end{equation}
so that $G(C)=G(0)$ in the incoherent regime. We derive from (\ref{wll}) in
the incoherent and coherent cases, respectively
\begin{eqnarray}
V(C) &=&V(0)\left| D/w(C)\right| ^{\rho G(0)}  \label{v1} \\
V(C) &=&V(0)/[1+G(0)\rho \ln |w(C)/D|]  \label{v2}
\end{eqnarray}
where $w(C)=C-E_f(C),$ $D$ is a cutoff parameter of the order of bandwidth
(we put in numerical calculations $D=\rho^{-1}=1$). Then we have the closed
equation for $w(C).$ In particular, for the incoherent regime

\begin{equation}
\frac{\partial w(C)}{\partial C}=1+(N-1)\frac{\rho V^2(0)}C\left| \frac
D{w(C)}\right| ^{2\rho G(0)}  \label{fcc}
\end{equation}

When $E_f$ lies sufficiently below the Fermi level (the Kondo regime), the
quantity $|w(C)|$ can become small with decreasing $|C|.$ We can use this
condition to define the boundary between IV and Kondo cases. Formal
definition of IV systems is the absence of solutions to the equation $w(C)=0$
which just determines the Kondo resonance (cf. Refs. %
\onlinecite{hewson,largeN}). Physically, Kondo lattice has a three-peak
density of states (two Hubbard bands and the Kondo resonance), which is
similar to the ``doped Mott insulator'' (note that in the dynamical
mean-field theory (DMFT) the Hubbard model is reduced to the Anderson
impurity model \cite{georges}). On the other hand, IV state is similar to
the phase of strongly correlated metal: the Kondo peak as a separate
solution is absent.

For $G(0)=0,$ the boundary condition for the Kondo state is $|E_f(0)|>\Gamma
=(N-1)\rho V^2(0).$ In the opposite IV case $|w(C)|$ remains finite. For $%
G(0)\neq 0$, $V(C)$ increases during the renormalization process, and the
effective level width $\Gamma (C)$ becomes larger, so that the IV region
becomes more wide. A temperature dependence of the energy gap (an increase
with decreasing temperature) in IV compounds was observed experimentally in
SmB$_6$ \cite{haliullin}, YbB$_{12}$ and Ce$_4$BiPt$_3$ \cite{rise2}.
According to our treatment, the dependence of the effective hybridization $\
V(C)$ is non-monotonous: it passes through a maximum.

Now we consider in more detail the incoherent case which should be realized
for diluted systems (the Anderson's localization prevents coherence at low
temperatures). To present numerical results (Figs.1-4) we use the variable $%
\xi =\ln |D/C|.$ As follows from (\ref{v1}), in the Kondo case the
hybridization parameter $V(C)$ decreases practically by a jump when we
approach the point $C=E_f(C),$ see Fig.2. With further decreasing $|C|$, a
considerable region arises where we have to high accuracy $C\cong E_f(C)$
(Fig.1). More exactly, we have $\partial w(C)/\partial C\simeq 0,$ so that
we obtain from (\ref{fcc}) near the maximum of $w(C)$%
\begin{equation}
w(C)\simeq -D\left| (N-1)\rho V^2(0)/C\right| ^{1/2\rho G(0)}.
\end{equation}
In this regime $V(C)=\left| C/[(N-1)\rho ]\right| ^{1/2},$ and the effective
$s-f$ exchange parameter is $\rho I(C)=\rho V^2(C)/E_f(C)=-1/(N-1)=\mathrm{%
const}$.

\begin{figure}[tbp]
\includegraphics[clip]{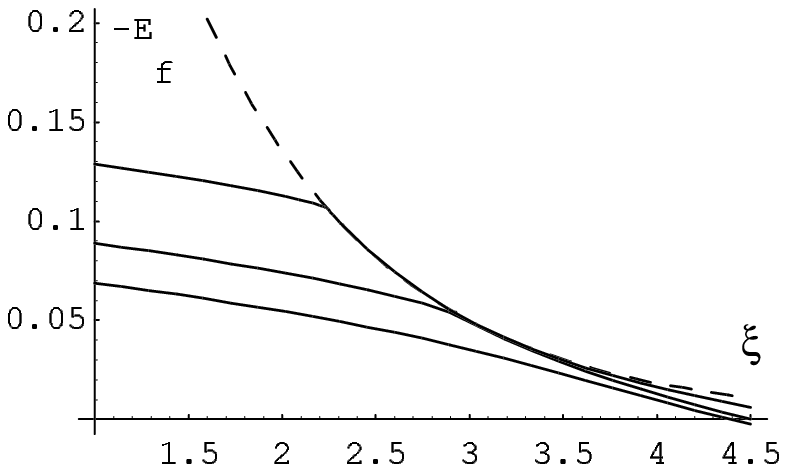}
\caption{Scaling trajectories $-E_f(\protect\xi )$ for $%
V(0)=0.1,G(0)=0.1,N=2 $ in the incoherent case as compared to the curve $%
|C|/D=\exp (-\protect\xi ) $ (dashed line). The parameter values (for the
curves from below to above) are $E_f(0)=-0.08,-0.1,-0.14.$}
\label{fig:1}
\end{figure}

\begin{figure}[tbp]
\includegraphics[clip]{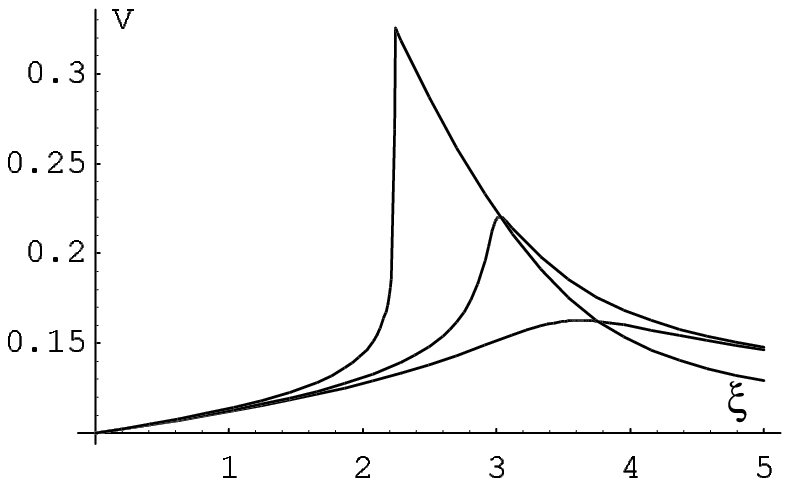}
\caption{Dependences $V(\protect\xi )$ for the same parameter
values as in Fig.1 (for the curves from below to above)}
\label{fig:2}
\end{figure}

\begin{figure}[tbp]
\includegraphics[clip]{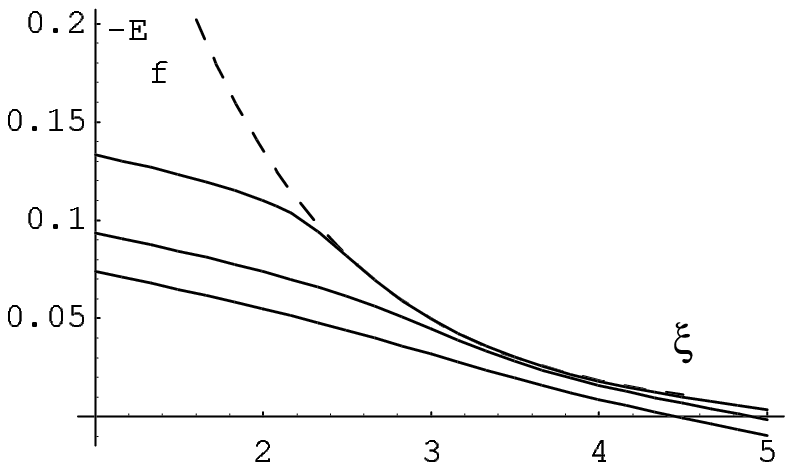}
\caption{Scaling trajectories $-E_f(\protect\xi )$ for $%
V(0)=0.05,G(0)=0.05,N=6,E_f(0)=-0.09,-0.11,-0.15$ in the coherent case. }
\label{fig:3}
\end{figure}

\begin{figure}[tbp]
\includegraphics[clip]{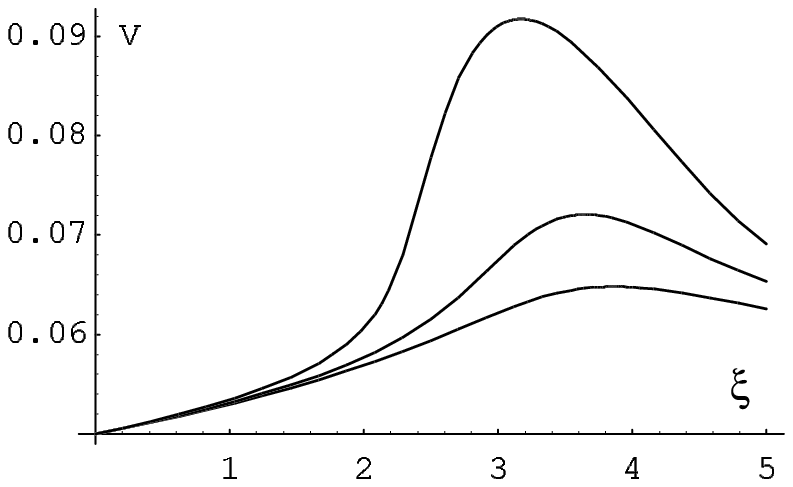}
\caption{Dependences $V(\protect\xi )$ for the same parameter
values as in Fig.3 (for the curves from below to above)}
\label{fig:4}
\end{figure}

In a standard consideration, the condition $C=E_f(C)$ determines an energy
scale for a crossover to the regime of a heavy-fermion (Kondo) local Fermi
liquid \cite{haldane}. The ``marginal'' situation with $E_f(C)\cong C$ in a
whole interval of the cutoff parameter $C$ means an essentially
non-Fermi-liquid (NFL) picture. A similar mechanism of NFL behavior in
magnetic Kondo lattices was proposed in Refs.\onlinecite{IK97,IKnfl} where a
soft boson was obtained with the characteristic energy $\overline{\omega }%
(C)\cong |C|$. Note also that a regime with the rate of the order parameter
fluctuations $1/\tau _\phi (T)\propto T$ is typical near a quantum phase
transition \cite{sach}. Our situation is reminiscent of this regime in the
sense that the characteristic valence fluctuation frequency $\max (\rho
V^2(T),|E_f(T)|)$ is proportional to $T$ (after a natural replacement $%
|C|\rightarrow T$).

Physically, the regime where a typical energy scale is just the temperature
means a \textit{classical} (Maxwell-Boltzmann) electron gas (interacting
with local momenta): heat capacity is approximately constant etc. Of course,
in a pure form this behavior is never observed since, according to our
results, it takes place only in a restricted temperature interval. However,
we have a strong deviation from a simple scaling picture where we just enter
strong-coupling regime, a characteristic ``Kondo'' temperature being $T$%
-independent. In particular, the Wilson number is nearly constant, but
differs considerably from that in the singlet state ($I_{ef}=-\infty $).

A similar ``marginal'' region in the dependence $E_f(C)$ occurs not only for
the Kondo, but also for IV state near the critical line (in such a
situation, the dependence $w(C)$ has a shallow minimum, and $V(C)$ a sharp
maximum, see Figs.1-2).

To estimate the Kondo temperature we can use Haldane's arguments for the
Anderson model with $N=2$ \cite{haldane}. The generalization to arbitrary $N$
can be performed as (cf. Ref. \onlinecite{rasul})
\begin{equation}
T_{K}\simeq D\left( \rho |I|N\right) ^{1/N}\exp \left( 1/\rho NI\right)
,I=V^{2}/E_{f}
\end{equation}%
This expression is formally based on perturbation theory (two-loop scaling).
However, Haldane noted that replacing in this formula both $D$ and $E_{f}$
by the characteristic energy scale $T^{\ast }=-C,$ which is determined from
the equation $C=E_{f}(C)$, yields the correct estimation for the Kondo
temperature. As demonstrated above, in the presence of exciton effects
(Falicov-Kimball interaction $G$), a situation is possible where this
equation holds approximately in a whole energy interval. The energy scale $%
T^{\ast }$ where the \textquotedblleft marginal\textquotedblright\ regime
starts is considerably changed by the exciton effects.

In the coherent case the last term in Eq.(\ref{gll}) results in a smearing
of the singularity, especially for small $N.$ However, with increasing $N$
the dependence $E_{f}(C)$ (Fig.3) becomes qualitatively similar to that in
the incoherent case. On the other hand, the dependence $V(C)$ (Fig.4) is
essentially modified even for $N\rightarrow \infty .$ At $C\cong E_{f}(C)$
we have
\begin{equation}
1=-(N-1)\rho V^{2}(C)/C-n[G(0)/V(0)]\partial V(C)/\partial C
\end{equation}%
Solution to this Riccati equation is obtained in terms of the
imaginary-argument Bessel and Macdonald functions $I_{p}(x)$ and $%
K_{p}(|(N-1)\rho C|^{1/2})$ with $\ x=2|(N-1)\rho
V^{2}(0)C|^{1/2}/[nG(0)],~p=0,1.$ For large $N,$ the quantity $\partial
V(C)/\partial C$ and consequently $V^{2}(\xi )/E_{f}(\xi )$ turn out to be
practically $\xi $-linear near the maximum of $w(C)$. Thus we have a
\textquotedblleft classical\textquotedblright\ electron liquid with singular
interactions which have logarithmic energy dependences. With further
decreasing $|C|$ we have from Eq.(\ref{wll})
\begin{equation}
\partial V(C)/\partial C\propto |C|^{-1/2}=|E_{f}(C)|^{-1/2}
\end{equation}%
so that the correction to $V^{2}(C)/E_{f}(C)$ is proportional to $%
|C|^{-1/2}. $

The $d-f$ Coulomb interaction can strongly renormalize the hybridization,
which leads to the increase of the characteristic energy scale. In the
incoherent regime it is a width of the resonance, in the coherent one the
width of (indirect) gap or pseudogap \cite{IK86}. The renormalization of the
fluctuation rate $\rho V^2(C)$ can be very strong (about order of magnitude
for realistic parameters). The corresponding temperature dependences can be
found in both the regimes by the RG approach with the replacement $%
|C|\rightarrow T$ .

In standard treatments of HF systems, one picks usually anomalous magnetic
contributions to thermodynamic properties and compares them with exact
results in the one-impurity Kondo problem. In particular, the dependences of
the crystal-field level width from both ineleastic neutron scattering and
nuclear magnetic resonance (NMR) have the form $\Gamma (T)\propto T^{1/2}.$
In the Kondo resonance model such a behavior takes place above the Kondo
temperature $T_{K},$ but in some cases (CeB$_{6}$, CePd$_{3}$B$_{0.6}$, YbBe$%
_{13}$) the dependence $T^{1/2}$ takes place at very low temperatures of a
few K \cite{nmr}. Further, the characteristic energy scale from linear heat
capacity term $\gamma $ is of order of tens of Kelvins, whereas the
temperature where $\gamma $ starts to deviate from constant is just a few of
Kelvins. Thus there exists no unique energy scale. We have demonstrated
that, indeed, the infrared behavior can be essentially different from the
simple Anderson model owing to spin dynamics (see Refs.%
\onlinecite{IK97,IKnfl}) and charge fluctuations (exciton effects)
considered in the present work.

Of course, the estimations performed are qualitative since they are based on
the continuation of perturbative Gellmann-Low scaling function to the strong
coupling region. At the same time, the statement that exciton effects cannot
be described by universal temperature independent $T_{K}$ seems to be
reliable itself. Recently, direct ways of observing the Kondo resonance
(STM) were proposed \cite{STM}. As we know, they have been not yet applied
to the IV systems. It would be interesting to compare the results for $T_{K}$
of these new experimental methods with those from investigating
thermodynamic properties.

The research described was supported in part by Grant No. 747.2003.2 from
the Russian Basic Research Foundation (Support of Scientific Schools).

\end{document}